\documentclass[preprint]{aastex}

\newcommand{\etal}{et al.\ }
\newcommand{\eg}{e.g., }
\newcommand{\ie}{i.e., }

\shorttitle{Testing MACHO Lens Candidates}
\shortauthors{von Hippel \etal}

\begin{document}

\title{Using Optical and Near-IR Photometry to Test MACHO Lens Candidates
\footnote{Based on data obtained at ESO (VLT/UT1) under project number
66.B-0326.}}

\author{Ted von Hippel}
\affil{The University of Texas at Austin, Department of Astronomy, 
1 University Station C1400, Austin, TX 78712-0259, USA \\
email: ted@astro.as.utexas.edu}

\author{Ata Sarajedini}
\affil{Department of Astronomy, 211 Bryant Space Science Center, P.O. Box
112055, University of Florida, Gainesville, FL 32611, USA \\
email: ata@astro.ufl.edu}

\author{Maria Teresa Ruiz}
\affil{Departamento de Astronomia, Universidad de Chile, Casilla 36-D, Santiago, Chile \\
email: mtruiz@das.uchile.cl}

\begin{abstract}

We obtained new VLT/ISAAC $H$-band observations for five MACHO LMC source
stars and adjacent LMC field regions.  After combining our near-IR
photometry with HST/PC $BVRI$ optical photometry, we compared the MACHO
objects to the adjacent field stars in a variety of color-magnitude and
color-color diagrams.  These diagnostic diagrams were chosen to be
sensitive to our hypothesis that at least some of the MACHO lenses were
foreground Galactic disk or thick disk M dwarfs.  For the five lensed
objects we studied, our hypothesis could be ruled out for main sequence
lens masses $\ga$ 0.1 M$_\sun$ for distances out to 4 kpc.  On the other
hand, the fact that LMC-MACHO-5, an object not in our study, has been
recently found to have just such a foreground lens, highlights that the
remainder of the LMC MACHO objects should be searched for the signature of
their lenses using our photometric technique, or via near-IR
spectroscopy.  We also constructed diagnostic color-color diagrams
sensitive to determining reddening for the individual MACHO source stars
and found that these five objects did not show evidence for significant
additional reddening.  At least these five MACHO objects are thus also
inconsistent with the LMC self-lensing hypothesis.

\end{abstract}


\section{Introduction}

Searches for gravitational microlensing along the line of sight toward the
Large Magellanic Cloud (\eg Paczynski 1986; Udalski, Kubiak, \& Szymanski
1997; Alcock \etal 2000a) were meant to test for the existence of Galactic
dark matter in the form of Massive Compact Halo Objects (MACHOs).  A
handful of gravitational microlensing events toward the LMC with the
duration and amplification expected for MACHOs have now been found.
Fitting these events into the dark matter picture or into the
well-constrained picture of Galactic structure has been problematic,
however.  For instance, the MACHO collaboration (Alcock \etal 2000a)
concluded that they had found 13 to 17 true microlensing events.  After
carefully modeling these events and comparing them to known sources of
photometric variability that might mimic microlensing, as well as known
stellar populations along the line of sight to the LMC, Alcock \etal
argued that they had detected a previously unknown Galactic halo
population of MACHOs with masses between 0.1 and 1 M$_\sun$.  Assuming the
assignment of these objects to the Galactic halo, these MACHOs are
abundant enough to account for approximately 20\% of the Galactic dark
matter implied by dynamical studies.  While the implied MACHO mass density
is dynamically significant, explaining one fifth of a problem with no hint
regarding the other four fifths of the problem is deeply unsatisfying.
More concretely, a large number of studies have found difficulties with
any suggested form of MACHO consistent with the mass range and other
properties.  For example, the CNO abundance patterns of Galactic stars
essentially rule out (\eg Gibson \& Mould 1997) the most discussed MACHO
candidate, cool white dwarfs.  Additionally, a halo of faint M stars is
inconsistent with the Hubble Deep Field observations (Gould \etal 1997).
If other galaxies contain halo MACHOs in the form of either white dwarfs
or M stars, these galaxy halos would be readily observable in the near-IR
(Charlot \& Silk 1995).  Given these difficulties, is there an alternative
explanation for the observed microlensing?  Some of the events seem to be
caused by background supernovae mimicing the expected microlensing light
curves (Alcock \etal 2000a), while other events are likely caused by LMC
self-lensing.  While these photometric variables may account for some of
the MACHO sources, we sought to test an additional form of contamination
among the MACHO events due to true gravitational microlensing, but by a
known and possibly underestimated population: M stars from the old disk
and thick disk.

The number of Galactic old disk and thick disk stars that contribute to
the lensing optical depth can be estimated from the local normalizations
and scaleheights of these populations, as well as their mass functions
down to approximately 0.1 M$_\sun$.  Unfortunately, all these parameters
except the old disk normalization and scaleheight are uncertain.  For
example, one of the most recent studies to count the faintest, lowest mass
Galactic stars (Gould \etal 1997) relied on a color - absolute luminosity
relation (Reid 1991) to determine stellar distances and thereby
densities.  If, however, the stars detected by Gould \etal are subluminous
compared to the local sample used by Reid (1991), then the implied density
would be much higher.  The time scale of lensing also is not a meaningful
discriminant between lensing by halo and thick disk objects, as the mean
values (101 days versus 129 days, Alcock \etal 2000a) differ by too little
with the current small samples.

If the MACHO lenses were garden variety old or thick disk M stars, there
would be clear observable consequences.  Such lenses still lie along the
line of sight to the LMC sources and would be observable as significant
additional near-IR light, measurable as an excess in the near-IR flux of
the LMC source.  We can check for this near-IR excess by comparing, for
example, $H$-band photometry to optical photometry for the MACHO objects.
Also, in a matter of time (see below) foreground disk or thick disk lenses
will move through a large enough angle to be detectable as separate
objects.  Note that our hypothesis is related to, but not identical to,
that of Gates \etal (1998), who argued for the disk as a partial source of
dark MACHO lenses.

Assuming our hypothesis is correct, the primary reason we should be able
to observe the lens as a near-IR excess is that we expect it to be
significantly closer than the LMC, \ie it should have a distance modulus
of $\sim$ 10 to 13, versus 18.5 for the LMC.  (Disk star lenses should be
distributed according to the combination of population scaleheight and
volume element, which gives a modal distance at two times the scaleheight,
\ie $\sim2$ $\times$ csc(b) $\times$ 350 pc and 2 $\times$ csc(b) $\times$
1000 pc for the old and thick disk, respectively.  For our viewing angle
through the Galactic disk toward the LMC, csc(b) = csc(33$^{\circ}$) =
1.8, so modal disk and thick disk stars will be found at 1.3 and 3.7 kpc,
respectively.)  This example also illustrates why we chose to perform this
experiment for the LMC sources rather than for the Galactic bulge lensing
sources, since the latter have a much smaller distance modulus,
approximately 14.5.  In addition, the nature of stellar populations along
the line of sight to the Galactic bulge is significantly more complex,
including the effects of large columns of dust and the Galactic bar.

We decided to search for near-IR excesses only among those lensing sources
which are LMC main sequence stars or slightly evolved subgiants, reasoning
that the LMC red giant lensing sources would be too red to notice any
near-IR contribution from foreground, line-of-sight Galactic M dwarf
lenses.  In retrospect, it would have been worthwhile to test our
hypothesis against some of the reddest LMC sources as well, and we return
to this point later.

\section{Data}

The MACHO collaboration obtained optical photometry with the Planetary
Camera on HST to image a number of the LMC lensing sources, in particular
to rule out background supernovae as MACHO mimics.  These HST data provide
the bulk of what we need for our experiment.  We selected a sample of six
MACHO sources with HST/WFPC2 multi-band photometry and locations in
optical color-magnitude diagrams indicative of being main sequence or
subgiant stars.  In addition, all but one of these objects (event \# 9)
pass the stricter lensing candidate selection criteria A of Alcock \etal
(2000a); criteria published after we selected our observing sample.  Our
approach is to supplement the HST optical photometry with near-IR
photometry.  The combination of optical and near-IR photometry allows us
to explore M dwarfs as possible lenses with masses as low as 0.1 M$_\sun$
out to 4 kpc, or higher mass M dwarfs to greater distances.

\subsection{VLT H-band Data}

We obtained $H$-band queue observations at VLT UT1 using ISAAC (Moorwood
2000) on November 6/7, 2000.  The VLT + ISAAC along with scheduling
observations in the queue allowed us to obtain data that were both
photometric and had good image quality--a challenge at the far southern
declination (Dec $\approx -70^{\circ}$) of the LMC, but necessary for the
relatively crowded LMC fields.  We obtained observations for 6 fields with
exposure times ranging from 390 to 780 sec and delivered image quality
ranging from 2.8 to 4.8 pixels, or 0.42 to 0.71 arcseconds (see Table 1
for details of the observations), along with sky fields taken near the
midpoint of the observing sequence.  Because of an error (by the PI, not
the VLT staff) in the phase II proposal process, one pointing missed the
target LMC source, and this field is dropped from further consideration.
The $H$-band was chosen as the best near-IR compromise between exposure
time to a depth of interest and sensitivity to the near-IR excess expected
for any line-of-sight M dwarfs.

The ISAAC data were reduced essentially as described in the ISAAC User's
Manual\footnote{http://www.eso.org/instruments/isaac/}.  The data were
dark subtracted, flat fielded, sky subtracted, residual bias corrected,
bad pixels were found and masked, and then all frames of a given field
were registered and averaged.  Some image irregularities remained, though
these were $\leq$ 0.5\%, and at least a fraction of this effect should be
additive and thus come out with the sky subtraction during the PSF-fitting
photometry.  Given the small range in airmass values for each field,
stacking all images of a given field should not compromise the
photometry.  For object identification we used SExtractor (Bertin \&
Arnouts 1996).  Since all five MACHO source objects were at least
partially blended with nearby stars, as were the majority of LMC field
stars, we employed DAOPHOT II/ALLSTAR (Stetson 1987) to fit empirical
point spread functions to the reduced and combined $H$-band data.  Between
50 and 100 bright uncrowded stars were used to construct the spatially
variable PSF on each image.  The ALLSTAR routine then iteratively fit the
PSF to the central regions of the detected profiles and calculated total
magnitudes by integrating the PSFs over their volumes.  Using the same 50
to 100 bright uncrowded stars, we then measured and applied a small,
spatially variable aperture correction to the PSF magnitudes to arrive at
total aperture magnitudes.

The photometry was calibrated using three standard stars (Persson \etal
1998) observed along with our program, although these standards were
observed at lower airmass values (X = 1.024--1.134) than our program
fields.  Since the number of standards observed was too few for a color or
airmass term determination, we used the mean $H$-band extinction value for
Paranal, 0.06 mag, and assumed a color term of 0.0, as indicated in the
ISAAC Data Reduction Guide (Amico \etal 2002).  Amico \etal expect the
$H$-band color term, using $J-K$ for the color, to be $\leq$ 0.01, but
have not yet measured its value.  We obtained an $H$-band zero point of
24.69 $\pm$ 0.02.  The zero point includes aperture corrections of $-0.015
\pm 0.01$ mag.  Uncertainties in the airmass term of order 0.01 to 0.02
mag airmass$^{-1}$ would cause $H$-band offsets of $\sim$ 0.006 to 0.012
mag in the extrapolation of the calibration to X $\sim$ 1.6 for all stars
in the field, whereas a small color term would cause a color-dependent
$H$-band offset.  While we have no way to estimate the error in the
assumed zero-valued color term, if a color term does exist for the
$H$-band it would not affect our primary goal of identifying an $H$-band
excess for the MACHO sources, since such an excess will be detected
relative to the other objects in the field with similar colors.

\subsection{HST/WFPC2 BVRI Data}

The MACHO team imaged a number of LMC microlensing sources, and these
observations are now available in the archive.  We used the CADC archive
to download the relevant archival images and recalibrate them with the
most up-to-date calibration files.  Since the HST data probed the LMC
fields to significantly greater depths than the $H$-band data, we did not
always use all of the available HST data when a subset in a particular
filter had a different roll angle or short exposures.  The properties of
the data we did use are listed in Table 2.  Since the MACHO sources were
centered in the PC chip, only the PC data were fully processed.  These
data were combined and cosmic rays rejected with the IRAF\footnote{IRAF is
distributed by the National Optical Astronomy Observatories, which are
operated by the Association of Universities for Research in Astronomy,
Inc., under cooperative agreement with the National Science Foundation.}
task CRREJ.  Sources were found with SExtractor, which also provided
morphological classification, allowing us to identify and reject
non-stellar objects.  The PC images, with a resolution of $\sim$ 2 pixels
or 0.092 arcseconds FWHM, were marginally crowded.  We were thus able to
employ aperture photometry, and chose CCDCAP\footnote{IRAF implementations
of CCDCAP are available via the web at the following site:~
http://www.noao.edu/staff/mighell/ccdcap/.} for its sub-pixel light
redistribution and accuracy with small apertures in marginally sampled HST
data (Mighell 1997).  An aperture of 4 pixels was used for all objects in
the PC fields, and aperture corrections to 0.5 arcseconds (11 pixels =
0.506 arcseconds) were determined empirically for the brighter objects in
each field and filter.  As is now well known, the WFPC2 CCDs suffer from a
time-dependent Charge Transfer Efficiency (CTE) problem.  These exposures
were short enough that the background counts on the PC frames ranged from
0 to $<$ 4 counts, and therefore CTE corrections were necessary,
especially for the fainter objects and objects at higher y-column values.
Time-dependent CTE corrections were installed based on the prescription of
Dolphin (2000) updated by the data he kindly makes available on his web
site\footnote{www.noao.edu/staff/dolphin/wfpc2\_calib/}.  The CCD dewar
window throughput also changes as a function of time and although such
changes are small for these filters (always $<$ 1.5\%), we applied these
corrections as well.  Finally, we applied the color corrections and zero
points to transform the F439W, F555W, F675W, and F814W instrumental
magnitudes to standard $B, V, R,$ and $I$ magnitudes using the
coefficients listed by Dolphin (2000) and updated on his web page.  The
photometric transformations of Dolphin are similar to those of Holtzman
\etal (1995) but updated for the CTE corrections one first applies, and
with some additional calibration information.  These calibrated HST data
should be on the $BVRI$ system of Landolt (1983, 1992) to within $\pm$
0.03 mag.

The optical PC data were merged with the requirement that an object be
detected in at least two filters.  The resulting merged photometry lists
were then merged with the $H$-band VLT data.  Transformations between the
PC and VLT reference frames were performed with great care, and typical
matches were possible within $\leq$ 0.3--0.5 ISAAC pixels, depending on
the field.  In a number of cases, besides the best matched object on the
PC frames, there were one or a few other PC detections within the $H$-band
FWHM.  During the final merging process we kept track of all such objects
and added their contributed light in each filter if the object would not
have been deblended (\ie if within 3 ISAAC pixels) by the $H$-band
PSF-fitting photometry.

Our photometry for the five MACHOs is presented in Table 3.  The optical
$V$ and $I$ magnitudes were previously presented by Alcock \etal (2001a).
We find a small difference between their and our results of $-0.09 \pm
0.08$ in $V$ and $-0.049 \pm 0.02$ in $I$, in the sense of their
photometry minus ours.  These differences are consistent with our use of
updated CTE corrections and photometric transformations.

Since most of the LMC MACHO sources were themselves blended and would not
have survived the blending cuts we apply to the remainder of the data
(below), we performed a final adjustment to their $H$-band photometry.  We
carefully examined the location, brightness, and optical colors of objects
in the region of each MACHO source on the HST/PC and VLT/ISAAC images to
determine, as best as possible, the likely contribution of any blended
objects that lie within the $H$-band PSFs.  We found probable
contributions by adjacent objects of $\sim$ 0.0, 0.2, 0.2, $\leq$ 0.1, and
0.05--0.1 mag in $H$ for LMC MACHO objects 4, 6, 8, 9, and 14,
respectively.  These expected $H$-band contributions are incorporated into
(removed from) the values presented in our figures and discussion, though
these corrections are {\it not} applied to the measured photometry listed
in Table 3.

\section{Discussion}

The $VI$ color-magnitude diagram (CMD) for all five LMC fields is
presented in Figure 1.  The five triangle symbols are the five MACHO
microlens source stars.  The adjacent field objects are also plotted if
their photometric errors were $<$ 0.1 mag, if any luminosity contributed
in the $I$-band from blended objects were $\geq$ 2.5 mag fainter than the
object of interest, and if their morphological classifications from
SExtractor were $\geq$ 0.9.  The blending criterion meant that objects
which had $\geq$ 10\% luminosity contributions from blended sources in $I$
were excluded.  Note that a typically blended object identified as a LMC
main sequence star will have a photometric companion that is also a main
sequence star.  The brightest such photometric (line-of-sight) companions
will have nearly the same color as the identified object, and thus will
add luminosity more effectively than optical color, much like the observed
binary sequences in clusters.  The situation is somewhat different for
colors composed of optical and $H$-band data, however, since the
ground-based $H$-band data had substantially larger PSFs, and thus objects
were generally not blended in the optical PC data but they may be blended
in the ISAAC $H$ data.  With regard to the morphological classification,
although the SExtractor classifier is not strictly Bayesian, the
morphology cut removes objects with $\la$ 90\% chance of being point
sources.  All five fields are overplotted since these LMC fields are only
a few degrees apart and therefore their distance moduli should be the same
within $\leq$ 0.03 mag (from equation 13 and the LMC geometry given by van
der Marel \& Cioni (2001) applied to these field positions).  Also plotted
are eight model sequences.  The left-most isochrone and the isochrone
approximately tracing the giant branch are [Fe/H] = $-0.7$ models for
log(age) = 8.5 and 9.5 (0.3 and 3.2 Gyr) populations, respectively.  We
use a 0.3 Gyr model to represent a zero age main sequence (ZAMS) and the
metallicity is chosen to approximate the mean observed value (\eg see
Cole, Smecker-Hane, \& Gallagher 2000) for the LMC.  The dotted line most
visible along the upper main sequence is the expected location of a solar
metallicity log(age) = 8.5 (ZAMS) population.  The dashed lines from left
to right represent the addition to the [Fe/H] = $-0.7$ young isochrone of
a 0.1, 0.2, 0.3, 0.4, and 0.5 M$_{\sun}$ star along the line of sight at a
distance of 4 kpc.  All models assume a distance modulus to the LMC of
18.5, consistent with Benedict \etal (2002), and E($B-V$)=0.1, consistent
with Harris, Zaritsky, \& Thompson (1997) who found E($B-V$)=0.13.  We
derived the stellar model loci from synthetic photometry of Lejeune,
Cuisinier, \& Buser (1997) model atmospheres appropriate for each
metallicity, effective temperature, and surface gravity for specific
Padova (Bertelli \etal 1994) isochrones.  The contribution by possible
line-of-sight low mass stars is necessarily approximate, since the
mass-luminosity relation for low mass stars is poorly known.  We used the
mass-luminosity relation of Henry \& McCarthy (1993) supplemented by
interpolation of the low mass main sequence data presented by Leggett
(1992).  The line-of-sight stars are assumed to be approximately solar
metallicity, as appropriate for Galactic disk stars.  Thick disk or halo
stars with lower metallicity would be brighter in both the optical and $H$
bands, and so more detectable at any give distance than solar metallicity
stars.  Our assumption of solar metallicity for any line-of-sight star is
thus conservative in terms of detectability.

Figure 1 reveals LMC field stars with an extended main sequence as well as
a subgiant branch, giant branch, and a red clump.  The age range of these
stars covers at least 0.3 to 3 Gyr.  The location of individual stars is
determined by their abundance and age, and possibly also by differential
reddening, image blending with other LMC stars, and line-of-sight low mass
star lenses for which we are searching.  Two of the MACHO sources appear
to be on or near the main sequence, two appear somewhat redder and
possibly consistent with a 0.2--0.3 M$_\sun$ main sequence star along the
line of sight at 4 kpc, and the remaining object could have a similar
explanation or could just be at the base of the giant branch.  The $VI$
CMD offers too little leverage to test for 0.1 M$_\sun$ line-of-sight
contributors, as can be seen from the dashed line for the 0.1 M$_\sun$
contributor that is minimally offset from the [Fe/H] = $-0.7$ isochrone.

Figure 2 presents the same models and data in the optical+near-IR $VH$ CMD
with the $H$-band error cut set at $<$ 0.2 mag, rather than at $<$ 0.1
mag, as is used for the optical bands.  Although the models and data now
cover a greater color range, this diagram does not break the degeneracy
between reddening, evolution away from the main sequence, and a
line-of-sight contribution to redness for the MACHO lenses.  Figure 3
presents a subset of these data and models in the $VRI$ color-color
diagram.  For clarity the 0.3 and 0.4 M$_{\sun}$ line-of-sight model
sequences have not been plotted and the photometric precision requirements
have been tightened to $\leq$ 0.02 mag in each band.  The 1 $\sigma$
photometric errors for the MACHO sources are smaller than the plotting
symbols.  This color-color diagram is independent of distance and
insensitive to reddening, as can be seen by the nearly parallel reddening
vector.  Likewise unimportant is metallicity; the solar and [Fe/H] =
$-0.7$ isochrones are nearly on top of each other.  Giants and dwarfs are
included in these sequences, and they too lie nearly on top of each
other.  Except for the bluest object, the MACHO sources all lie among the
sequence delineated by the LMC field stars and are consistent with no
line-of-sight low mass star lens, although this color-color diagram is not
meaningfully sensitive to lenses with mass as low as 0.1 M$_{\sun}$ at 4
kpc.  Figure 4 presents the LMC field stars in the $VRH$ color-color
diagram, along with the same model stellar sequences.  Because of the wide
range of the $V-H$ axis and the shallower limits of the $H$-band
photometry, the photometry error selection is relaxed to $<$ 0.2 mag for
the $H$-band.  The 1 $\sigma$ photometric errors for the five MACHO
sources range from about half the size to about one and a half times the
size of the plotting symbols.  The reddening vector is also largely
parallel to the stellar sequence in Figure 4 and metallicity has again
essentially no effect in this color-color diagram.  Now, with greater
sensitivity to any low mass main sequence lenses, the MACHO source
photometry for all but the bluest object is inconsistent with low mass,
lensing main sequence stars, unless they have masses less than 0.1
M$_{\sun}$ or distances $>$ 4 kpc.

With photometry through five filters ($BVRIH$) available to us for four of
our five fields and photometry through four filters ($VRIH$) available to
us for the remaining (LMC-MACHO-14) field, we were able to create a large
number of CMDs and color-color diagrams.  We do not present these diagrams
here as they add little extra insight.  The optical CMDs and color-color
diagrams all appear to be variations on the themes of Figures 1 and 3, and
the $H$-band CMDs and color-color diagrams appears to be variations on
Figures 2 and 4.

Before proceeding on to further interpretation we verify the location of
the model main sequences.  In Figure 5 we compare the location of the
models to modern photometry for open clusters in the $VRH$ color-color
diagram.  The open cluster data are a combination of optical photometry
for M35 (Deliyannis \etal 2003), M67 (Montgomery, Marschall, \& Janes
1993), and NGC 2420 (Anthony-Twarog \etal 1990) along with 2MASS $H$-band
photometry merged with the optical data by Grocholski \& Sarajedini
(2003).  The fit is excellent, with only a $\Delta$($V-H$) $\approx$ 0.1
mag offset for a given $V-R$ color out to $V-H \approx$ 2.5.  The CMD fits
are also excellent, though there one has the adjustable parameters of
distance, metallicity, and reddening.

\subsection{Blending}

With 800--2500 stars per LMC HST/PC field down to $I \approx$ 24, or
0.6--1.8 stars per square arcsecond, the degree of blending seen in the
$H$-band data is expected.  For the HST/PC with an image quality of $\sim$
2 pixels and nearly 500 pixels per square arcsecond, blending is typically
not a problem.  The PC data thus provides an excellent indicator for the
expected blending in the ISAAC $H$-band data.  We used the $I$-band data
for the blending measurement since these data are closest in wavelength to
the $H$-band data, and since the $I$-band reaches about 4 mag deeper than
the $H$-band data.  Thus, for any $H$-band detections more than 1 mag
above the $H$-band limit we are able to determine whether there is any
blending contribution to $\leq$ 1\%.  In the preceding analysis and plots,
we have however presented objects with $I$-band blending as large as
10\%.  Could this $\leq$ 10\% blending, perhaps by intrinsically redder
stars which would contribute relatively more in $H$ than in $I$, have
caused problems in the interpretation?  In Figure 6 we present the $VRH$
color-color diagram with a blending limit of $\leq$ 1\%.  The results are
the same, though there are fewer LMC stars to see the general trend.  This
more stringent blending cut was tested in all CMDs and color-color
diagrams, with no net effect on the general location of the LMC field
stars.

\subsection{LMC Self-Lensing}

The optical depth of LMC self-lensing has been examined in detail by a
number of authors (\eg Sahu 1994; Gyuk, Dalal, \& Griest 2000; Zhao
2000).  In this section we examine whether our photometry can constrain
the contribution of LMC self-lensing.  The addition of the $H$-band to the
optical data allows us to construct a variety of reddening-free versus
reddening sensitive color-color diagrams.  Reddening-free color
combinations can be created from any two color combinations linearly
differenced in proportion to their reddening ratio.  In classic Johnson
$UBV$ photometry these reddening free parameters are abbreviated by $Q$
(\eg see Mihalas \& Binney 1980).  We constructed eight reddening free
indices, which we shall refer to as $Q_1$ through $Q_8$, and plotted those
against the $V-H$ color, the longest wavelength baseline color we had for
every field.  The $V-H$ color is the most sensitive to reddening: E($V-H$)
= 2.58 E($B-V$).  The coefficients in the reddening relations are based on
equations 3a and 3b of Cardelli, Clayton, \& Mathis (1989).  The
hypothesis we are testing is that many of the lens sources are more
distant than the majority of the LMC field stars, which is a prediction of
the self-lensing hypothesis, and therefore on average they will be more
reddened (Zhao 2000).  In comparing the lensed stars with objects close to
them on the sky in a reddening sensitive versus reddening-free diagram,
the lensing sources should stand out as typically being more reddened.  We
initially chose as comparison objects all stars within 4 arcseconds of the
microlens source star, but subsequently relaxed the location criterion to
the full HST/PC field of view since there was no statistical difference in
the location of the field stars in this small area versus the entire PC
field and since including more objects make it easier to locate the MACHO
sources relative to the field LMC stars.  Note that reddening due to dust
along the line of sight is not the same sort of reddening we were studying
above, which would instead be caused by the addition of near-IR light from
a much redder object along the line of sight.  In various CMDs either type
of reddening might have the same effect, but they can in principle be
differentiated with the appropriate color-color diagrams.

Of the eight $Q$ versus $V-H$ diagrams we constructed, three turned out to
have the greatest diagnostic potential due to the slope of the stellar
sequence through the diagram.  These three diagrams are presented in
Figures 7a-c.  The data and model sequences presented in the earlier
figures are represented in these reddening-free versus reddening-sensitive
color-color diagrams.  The MACHO source stars are plotted with their 1
$\sigma$ error bars to distinguish them from the LMC field stars.  The
three MACHO source stars of intermediate $V-H$ color are consistent in two
out of three of these diagrams with the average reddening of the field.
In two out of three $Q$ diagrams the bluest object, near $V-H$ = 0.2, is
within the envelope of the data but otherwise consistent with a deficit in
reddening of $\sim0.8$ in $V-H$ or $\sim0.3$ in $B-V$.  Such a deficit
in reddening is too large given the small overall reddening, and is
aphysical.  All three $Q$ diagrams can be interpreted as implying excess
reddening for the reddest MACHO source star, or alternatively, given the
shape of the stellar sequences, two out of the three $Q$ diagrams are
consistent or marginally consistent with no excess reddening.  The large
amount of reddening required in the reddening interpretation for this
object would equal an additional E($V-H$) $\approx$ 1.5 or E($B-V$)
$\approx$ 0.6.  This is more than necessary to place this object behind
the LMC.  The most parsimonious explanation for the $Q$ diagrams of the
five MACHO source stars is that they have typical reddening for their
field and that the color variations seen represent another effect, for
instance photometric blending.  We conclude that for these five MACHO
sources there is no significant evidence that they are excessively
reddened, as one would expect if they had been self-lensed.

\subsection{Galactic Main Sequence Stars as Lenses}

Our experiment was specifically designed to test for the contribution of
thin and thick disk low mass dwarfs as MACHO lenses.  Our search was
sensitive to main sequence lensing objects with mass $\geq$ 0.1 M$_\sun$
out to distances of 4 kpc.  This statement has been generalized in Figure
8, where we show as a function of LMC source star mass the limiting, or
maximum, distance for detecting 0.1, 0.2, 0.3, 0.4, and 0.5 M$_\sun$ main
sequence lenses by our technique.  We assume that lensing objects which
add $V-H=0.1$ mag to the color of the source star would be detectable,
based on the limits found above.  We further assume solar metallicity
lenses with the mass-luminosity relation of Henry \& McCarthy (1993), and
the [Fe/H]=0.0, $-0.7$, or $-1.3$, 0.3 Gyr Bertelli \etal (1994)
isochrones and Lejeune \etal (1997) stellar atmospheres, as used above.
These models and assumptions are consistent with the previous analyses.
Lower metallicity main sequence lensing stars would be somewhat brighter
for a given mass, and therefore easier to detect via our technique.  At
present we do not know the metallicities for the LMC MACHO source stars,
but it is reasonable to assume that they have approximately the same
metallicity as the average star in the LMC.  Changes in source star
metallicity to solar and [Fe/H] = $-1.3$ are shown for the 0.5 and 0.1
M$_\sun$ lens cases.  The intermediate lens mass cases look similar and so
are not presented to avoid unnecessary crowding in the figure.  The region
below each line represents the parameter region in which our technique
would have detected main sequence lenses.  The horizontal dashed-dotted
line is placed at 4 kpc, a pathlength at the Galactic latitude of the LMC
of $\sim2$ $\times$ the scaleheight of the thick disk.  Four of the five
MACHO sources are most likely on the main sequence (see below) or only
slightly evolved.  Assuming their metallicity to be [Fe/H]=$-0.7$, their
masses are derived by interpolation in our synthetic photometry of the
Bertelli \etal (1994) isochrone and Lejeune \etal (1997) atmosphere
models.  These masses are indicated in Figure 8 by the four arrows.  Our
technique would have detected the MACHO lenses over a wide range of
distance and lens mass, assuming they were main sequence stars.  For three
of the MACHOs (\# 4, 8, and 6) such a lens can be ruled out for mass
$\geq$ 0.1 M$_\sun$, and for slightly higher lens masses even placing the
lens in the distant halo would not escape detection.  The situation for
MACHO \# 14 is less clear, as indicated in Figure 8 and Figure 4, where
this object with a slight $V-H$ excess may match a variety of
line-of-sight lens masses.  MACHO \# 9 cannot be so easily compared to an
isochrone since one star near the base of the giant branch can match a
wide variety of isochrones.  On the other hand, in the $VRH$ color-color
diagram, this object is too blue in $V-H$ for its $V-R$ color.  Since this
diagram collapses the main sequence, subgiant, and giant stars onto the
same color-color sequence, a near-IR excess is still expected for this
object if it had a main sequence lens, though this technique is less
sensitive due to the source star's redder color.  For the detailed
properties of this LMC star, a lensing main sequence star at 4 kpc with
mass = 0.1 or 0.2 M$_\sun$ would produce a $V-H$ excess of 0.06 or 0.19
mag, so the technique remains sensitive down to nearly 0.1 M$_\sun$.

\subsection{Details of Individual Lensing Sources}

The above analyses demonstrates that the optical/near-IR colors of the
five MACHO source stars studied here are minimally affected or even
entirely unaffected by their lenses.  The CMD for these stars can then be
interpreted in the standard fashion.  Object \#4 appears to be either a
main sequence star with [Fe/H] $>$ $-0.7$, or a turn-off star with [Fe/H]
$\approx -0.7$.  Assuming the models we have employed and [Fe/H] = $-0.7$,
its mass is 1.37 M$_\sun$.  Object \#6 is currently evolving away from the
main sequence.  For the same model assumptions its mass is 1.99 M$_\sun$.
Object \#8 has a mass of 1.78 M$_\sun$ and is also likewise somewhat
evolved off the main sequence.  Object \#9 is the reddest star and is
currently at the base of the giant branch.  Its main sequence mass depends
on the isochrone fit to this single object, but it is most likely slightly
more massive than object \#4.  The photometry for object \#9 is slightly
suspect since Alcock \etal (2001a) could not identify which of two
adjacent stars had been the lensing source.  Both of these objects are
fortuitously subgiants, however, so only a small photometric error
($\Delta V$ = 0.113, $\Delta V-I$ = 0.011) would result from assuming the
wrong object in the HST/PC frames.  Object \#9 also underwent a binary
lensing event, so may itself be a binary or it was lensed by a binary
(Alcock \etal 2000b).  Object \# 14 appears to be a minimally evolved main
sequence star with a mass of 2.37 M$_\sun$.

Interestingly, we identify faint nebulosity around objects \#8 and \#14 in
the HST/PC images.  Alcock \etal (2000a) concluded that none of these
objects coincided with background galaxies and thus were not supernovae.
This assumption should be re-examined for these two objects, though deeper
imaging may be required.

\subsection{LMC-MACHO Candidate 5}

One of the LMC MACHO candidates (\#5) that is not part of this study was
found to be behind a nearly line-of-sight disk M dwarf star in follow-up
HST/PC imaging and VLT spectroscopy  (Alcock \etal 2001b).  This M star
was separated from the photometric center of the source star by only 0.134
arcseconds.  Assuming it was the lens gives it a proper motion consistent
with the properties of the lens based on the MACHO light curve analysis.
The conclusion of Alcock \etal (2001b) was that finding one such object
among the LMC gravitational microlenses was entirely consistent with the
expected microlensing optical depth of the Galactic disk population.  Here
we take a brief detour, using this object as a guide, to ask whether our
technique would have discovered this object before it became visually
separated on the sky and how long one might have to wait for other such
lenses to move far enough away from the line of sight of their source
objects to be identified as a separate objects in HST/PC photometry.

From the properties listed for this M dwarf star by Alcock \etal (2001b),
we can add this object to the HST photometry of the LMC-MACHO 5
candidate.  Unfortunately, different techniques yield different masses for
the lensing object.  Their constrained lensing fit yields a mass of 0.036
M$_\sun$ and their direct lensing calculation yields a 2 $\sigma$ upper
limit mass of 0.069 M$_\sun$.  From the lens parameters they derive a
distance of 170--240 pc and $M_V$ = 15.7--16.8.  On the other hand, the
VLT/FORS2 spectrum for the candidate lens indicates that it is a M4-5
dwarf, consistent with the HST optical colors.  The mass for such an
object is significantly higher than their lensing estimates, at
0.095--0.13 M$_\sun$, depending on its metallicity.  From the spectrum
they derive $M_V$ = 13.61 $\pm$ 0.55 and d = 650 $\pm$ 190 pc.  We prefer
the latter interpretation under the assumption that the observed object is
the lens, since the mass and absolute magnitude estimate are based on the
observed spectrum and colors.  Such an object would be detected by our
technique, as is clear from the mass estimate near 0.1 M$_\sun$ and the
distance which is much less than 4 kpc.  Under the assumption that the
object is an M4V or M5V disk star it would have $V-H$ = 5.01--5.84
(Tokunaga 2000), and thus $M_H$ = 7.22-9.15, taking the extremes in the
1$\sigma$ errors in $M_V$ and the color estimate.  Using the observed
source and lens optical photometry of Alcock \etal (2001b), a $+1\sigma$
distance of 840 pc, the lower $H$ flux implied by $V-H$ = 5.01, and
assuming $V-H$(source) = 1.4, as we see in our LMC field stars for objects
with this source star's $V-I$ color, the expected properties of this
object in our CMDs and color-color diagrams are presented in Figures 1 to
4 with star symbols.  In each of these figures the source star alone is
plotted, as well as the photometrically merged object, which is the redder
object in all four of these diagrams.  If this apparent lensing object
were still photometrically merged with this MACHO source, its properties
would highlight it as an extremely red object in the CMDs.  The $VRI$
diagram would be the most indicative of the true properties of this
object, although the $VRH$ diagram would also indicate a likely foreground
low mass main sequence star lens.  Note that this analysis suggests that
the LMC source stars that are apparently redder than any expected stellar
sequence (objects \# 5, 11, 19, 20, and 24 in Alcock \etal 2000a) are
prime suspects for lensing by a foreground M dwarf.  Secondary suspects
are the other red objects (\# 1, 16, 17, 18, and 25).

The Einstein ring radius for an object at 4 kpc is $\sim5.9$ $\times$
sqrt(mass) milliarcseconds (see equation 16 of Paczynski 1986).  Assuming
this object has a mass of 0.1 M$_\sun$, the Einstein ring radius is
$\sim$1.9 mas.  This is 50 times smaller than the diameter (FWHM)
corresponding to the delivered image quality of the HST/PC data.  A
gravitational microlensing event lasting two months would correspond to a
photometric alignment within the PC image that lasted approximately 8
years, split evenly between time before and time after the lensing epoch.
Typical MACHO lenses should be emerging from photometric alignment with
their LMC source stars when viewed at HST/PC or HST/ACS resolution.  The
number of photometrically aligned objects that will not become microlenses
is even larger, approximately the square of the above radius ratio, or
$\sim$625 for the PC image quality.  The total MACHO inventory of stars,
from among which somewhat more than a dozen lenses were found, contains
$\sim10^7$ stars.  Approximately 0.01\% of these objects should have
photometrically aligned foreground stars at HST/PC resolution that will
not cause lensing.

\section{Conclusions}

We obtained new VLT/ISAAC $H$-band observations of a handful of MACHO LMC
source stars and adjacent LMC field regions.  After combining our new
near-IR photometry with $BVRI$ optical photometry rederived from HST/PC
imaging, we compared the MACHO objects to the adjacent field stars in a
variety of color-magnitude and color-color diagrams.  These diagnostic
diagrams were chosen to be sensitive to our hypothesis that at least some
of the MACHO lenses were foreground Galactic disk or thick disk M dwarfs.
For the five lensed objects we studied, our hypothesis could be ruled out
for main sequence lens masses $\ga$ 0.1 M$_\sun$ for distances out to 4
kpc.  On the other hand, the fact that LMC-MACHO-5, an object not in our
study, has been recently found (Alcock \etal 2001b) to have just such a
foreground lens, highlights that the remainder of the LMC MACHO objects
should be searched for the signature of their lenses using our photometric
technique, or via near-IR spectroscopy.  A number of the remaining LMC
MACHO objects are excessively red, based on their positions in the MACHO
CMD, and these in particular should be investigated.  We also constructed
diagnostic color-color diagrams sensitive to determining reddening for the
individual MACHO source stars and found that these five objects did not
show evidence for significant additional reddening.  At least these five
MACHO objects are thus also inconsistent with the self-lensing
hypothesis.

We also recommend an extension of our technique for future MACHO
searches: instead of complementing optical photometry with near-IR
photometry (or spectroscopy) after the lensing event has past, these
near-IR data could be obtained both during and after the lensing event.
Such lensing-event data would break the degeneracy between excess
reddening in the source star and the lensing object.

\acknowledgements

We thank the Canadian Astronomy Data Centre for the archival HST data with
recalibration and the VLT staff for acquiring the high quality queue
observations.  This publication makes use of data products from the Two
Micron All Sky Survey, which is a joint project of the University of
Massachusetts and the Infrared Processing and Analysis Center/California
Institute of Technology, funded by the National Aeronautics and Space
Administration and the National Science Foundation.  We also thank Andrew
Dolphin for helpful discussions on calibration and Cailen Nelson for MACHO
finding charts.  AS was supported by NSF CAREER grant AST-0094048.  MTR
received partial support from FONDAP (15010003), a Guggenheim Fellowship,
and Fondecyt (1010404).  TvH thanks McDonald Observatory for partial
support of this project.


\references

Alcock, C. \etal 2000, ApJ, 541, 270 (2000b)

Alcock, C. \etal 2000, ApJ, 542, 281 (2000a)

Alcock, C. \etal 2001, ApJ, 552, 582 (2001a)

Alcock, C. \etal 2001, Nature, 414, 617 (2001b)

Amico, P., Cuby, J. G., Devillard, N., Jung, Y., \& Lidman, C. 2002, ISAAC
Data Reduction Guide 1.5
(http://www.eso.org/instruments/isaac/drg/html/drg.html)

Anthony-Twarog, B. J., Twarog, B. A., Kaluzny, J., \& Shara, M. M. 1990,
AJ, 99, 1504

Benedict, G. F. \etal 2002, AJ, 123, 473

Bertelli, G., Bressan, A., Chiosi, C., Fagotto, F., \& Nasi, E. 1994,
A\&AS, 106, 275

Bertin, E., \& Arnouts, S. 1996, \aaps, 117, 393

Cardelli, J. A., Clayton, G. C., \& Mathis, J. S. 1989, ApJ, 345, 245

Charlot, S., \& Silk, J. 1995, ApJ, 445, 124

Cole, A. A., Smecker-Hane, T. A., \& Gallagher, J. S. 2000, AJ, 120, 1808

Deliyannis, C. P., \etal 2003, in preparation

Dolphin, A. E. 2000, PASP, 112, 1397

Gates, E. I., Gyuk, G., Holder, G. P., \& Turner, M. S. 1998, ApJ, 500,
L145

Gibson, B. K., \& Mould, J. R. 1997, ApJ, 482, 98

Gould, A., Bahcall, J. N., \& Flynn, C. 1997, ApJ, 482, 913

Grocholski, A., \& Sarajedini, A. 2003, in preparation

Gyuk, G., Dalal, N., \& Griest, K. 2000, ApJ, 535, 90

Harris, J., Zaritsky, D., \& Thompson, I. 1997, AJ, 114, 1933

Henry, T. J., \& McCarthy, D. W. 1993, AJ, 106, 773

Holtzman, J. A., Burrows, C. J., Casertano, S., Hester, J. J., Watson,
A. M., \& Worthy, G. S. 1995, PASP, 107, 1065

Landolt, A. U. 1983, AJ, 88, 439

Landolt, A. U. 1992, AJ, 104, 340

Leggett, S. K. 1992, ApJS, 82, 351

Lejeune, Th., Cuisinier, F., \& Buser, R. 1997, A\&AS, 125, 229

Mighell, K. J. 1997, AJ, 114, 1458

Mihalas, D., \& Binney, J. 1981, Galactic Astronomy: Structure and Kinematics,
(San Francisco: W. H. Freeman \& Co), 186

Montgomery, K. A., Marschall, L. A., \& Janes, K. A. 1993, AJ, 106, 181

Moorwood, A. 2000, in From Extrasolar Planets to Cosmology, ed.\ J.
Bergeron \& A. Renzini, (Berlin: Springer-Verlag), 12

Paczynski, B. 1986, ApJ, 304, 1

Persson, S. E., Murphy, D. C., Krzeminski, W., Roth, M., \& Rieke, M. J.
1998, AJ, 116, 2475

Reid, N. 1991, AJ, 102, 1428

Sahu, K. C. 1994, PASP, 106, 942

Tokunaga, A. T. 2000, Allen's Astrophysical Quantities, ed. A. N. Cox,
(Springer-Verlag: New York), p. 151

Udalski, A., Kubiak, M., \& Szymanski, M. 1997, AcA, 47, 319

van der Marel, R. P., \& Cioni, M.-R. L. 2001, AJ, 122, 1807

Zhao, H. S. 2000, ApJ, 530, 299

\clearpage

\figcaption{The LMC MACHO sources (triangle symbols) and LMC field stars
in the $VI$ CMD.  Models based on the Bertelli \etal (1994) isochrones and
the Lejeune \etal (1997) model atmospheres for age=0.3 Gyr with
[Fe/H]=$-0.7$ and [Fe/H]=0.0 are indicated by the left-most solid line and
the dotted line, respectively.  A 3.3 Gyr, [Fe/H]=$-0.7$ model is also
presented.  Models for a combination of the 0.3 Gyr, [Fe/H]=$-0.7$
sequence and 0.1, 0.2, 0.3, 0.4, and 0.5 M$_\sun$ line-of-sight main
sequence disk stars at 4 kpc are presented as dashed lines.  The star
symbols represent the expected locations for the source-only and combined
photometry for LMC-MACHO-5.}

\figcaption{The LMC MACHO sources and field stars in the $VH$ CMD.
Symbols have the same meaning as in Figure 1.}

\figcaption{The $VRI$ color-color diagram for the LMC data, ZAMS models,
and combination lens+source models.  For clarity the 0.3 and 0.4 M$_\sun$
lens models are not presented.  The reddening vector for E($B-V$)=0.1 is
also presented.}

\figcaption{The $VRH$ color-color diagram for the LMC data, ZAMS models,
and combination lens+source models.  Symbols have the same meaning as in
Figure 3.}

\figcaption{The $VRH$ color-color diagram for the models along with data
for the open clusters M35, M67, and NGC 2420.}

\figcaption{The $VRH$ color-color diagram for the LMC data, ZAMS models,
and combination lens+source models, but now the blending threshold has
been reduced from 2.5 mag to 5.0 mag.}

\figcaption{$Q$ versus $V-H$ diagrams for the LMC data and the 0.3 Gyr,
[Fe/H]=$-0.7$ and [Fe/H]=0.0 models.  $Q_4$ ({\it a}), $Q_5$ ({\it b}),
and $Q_7$ ({\it c}) are composed of different color-color combinations, as
listed on the vertical axes, which are insensitive to reddening.  To
identify the LMC MACHO source photometry from the field star photometry,
the MACHO sources include their $1\sigma$ error bars.}

\figcaption{The maximum detectable distances for main sequence lenses of
0.1, 0.2, 0.3, 0.4, and 0.5 M$_\sun$ as a function of the mass of the LMC
source star, assumed to have [Fe/H]=$-0.7$.  For the 0.1 and 0.5 M$_\sun$
case the dotted and dashed curves present the same quantities under the
assumptions of solar metallicity and [Fe/H]=$-1.3$ for the LMC source
star.  The arrows at the bottom of the figure indicate the expected masses
of the listed MACHO source stars and the dashed-dotted line indicates a
distance equivalent to twice the thick disk scaleheight for the Galactic
latitude of the LMC.}

%
\begin{deluxetable}{cccc}
\tablewidth{0pt}
\tablecaption{Log of VLT Observations}
\tablehead{
\colhead{object} & \colhead{exposure} & \colhead{DIQ}      & \colhead{airmass} \\
\colhead{}      & \colhead{(sec)}    & \colhead{(pixels)} & \colhead{}}
\startdata
 4 & 780 & 3.5--4.8 & 1.538--1.560 \\
 6 & 624 & 2.8--3.4 & 1.511--1.527 \\
 7 & 780 & 2.8--3.3 & 1.412--1.415 \\
 8 & 780 & 2.8--3.2 & 1.428--1.436 \\
 9 & 390 & 3.0      & 1.408--1.411 \\
14 & 390 & 3.1--3.7 & 1.600--1.614
\enddata
\end{deluxetable}

%
\begin{deluxetable}{cccc}
\tablewidth{0pt}
\tablecaption{Log of HST Observations}
\tablehead{
\colhead{object} & \colhead{filter} & \colhead{exposure} & \colhead{epoch}}
\startdata
 4      & F439W  &  1620    &  8-19-99 \\
\nodata & F555W  &  1590    & 12-12-97 \\
\nodata & F675W  &  1590    & 12-12-97 \\
\nodata & F814W  & 16500    & 11-12-97 \\
 6      & F439W  &  1620    &  8-26-99 \\
\nodata & F555W  &  1620    &  8-26-99 \\
\nodata & F675W  &   820    &  8-26-99 \\
\nodata & F814W  &  1040    &  8-26-99 \\
 8      & F439W  &  1600    &  3-12-99 \\
\nodata & F555W  &  1620    &  3-12-99 \\
\nodata & F675W  &   820    &  3-12-99 \\
\nodata & F814W  &  1020    &  3-12-99 \\
 9      & F439W  &  1600    &  4-13-99 \\
\nodata & F555W  &  1620    &  4-13-99 \\
\nodata & F675W  &   820    &  4-13-99 \\
\nodata & F814W  &  1020    &  4-13-99 \\
14      & F555W  &  2120    &  5-13-97 \\
\nodata & F675W  &  2120    &  5-13-97 \\
\nodata & F814W  &  2120    &  5-13-97
\enddata
\end{deluxetable}

%
\begin{deluxetable}{cccccc}
\tablewidth{0pt}
\tablecaption{Photometry of MACHO Sources}
\tablehead{
\colhead{object} & \colhead{B} & \colhead{V} & \colhead{R} & \colhead{I} & \colhead{H}}
\startdata
 4 & 21.73 $\pm$ 0.02 & 21.41 $\pm$ 0.01 & 21.17 $\pm$ 0.01 & 20.90 $\pm$ 0.01 & 20.28 $\pm$ 0.11 \\
 6 & 20.46 $\pm$ 0.01 & 20.07 $\pm$ 0.01 & 19.88 $\pm$ 0.01 & 19.62 $\pm$ 0.01 & 18.93 $\pm$ 0.03 \\
 8 & 20.60 $\pm$ 0.01 & 20.40 $\pm$ 0.01 & 20.22 $\pm$ 0.01 & 20.02 $\pm$ 0.01 & 19.46 $\pm$ 0.06 \\
 9 & 22.28 $\pm$ 0.03 & 21.39 $\pm$ 0.01 & 20.79 $\pm$ 0.01 & 20.38 $\pm$ 0.01 & 19.16 $\pm$ 0.05 \\
14 & \nodata          & 19.52 $\pm$ 0.01 & 19.51 $\pm$ 0.01 & 19.41 $\pm$ 0.01 & 19.25 $\pm$ 0.06
\enddata
\end{deluxetable}

\clearpage

\begin{figure}[]
\plotone{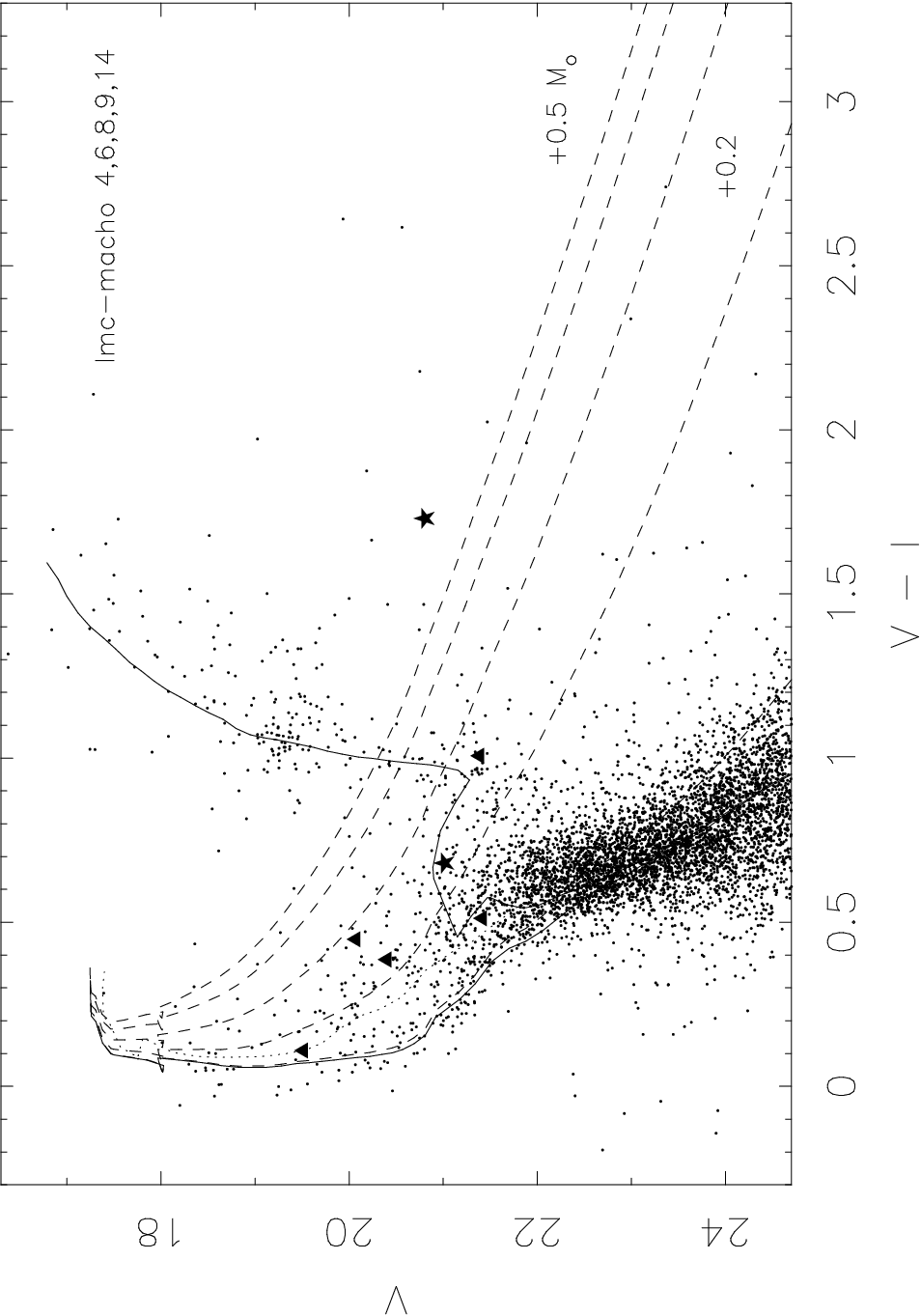}
\end{figure}

\begin{figure}[]
\plotone{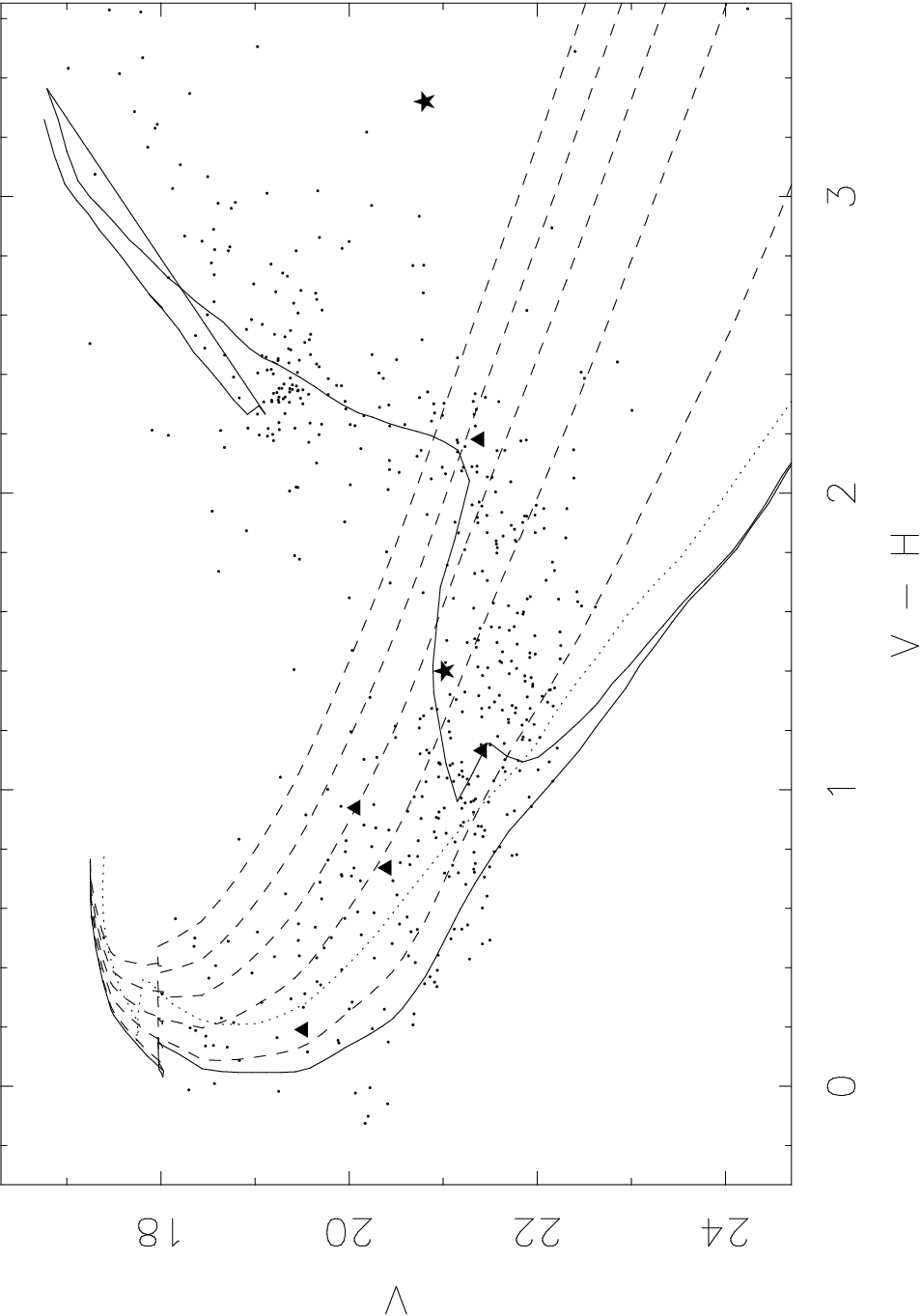}
\end{figure}

\begin{figure}[]
\plotone{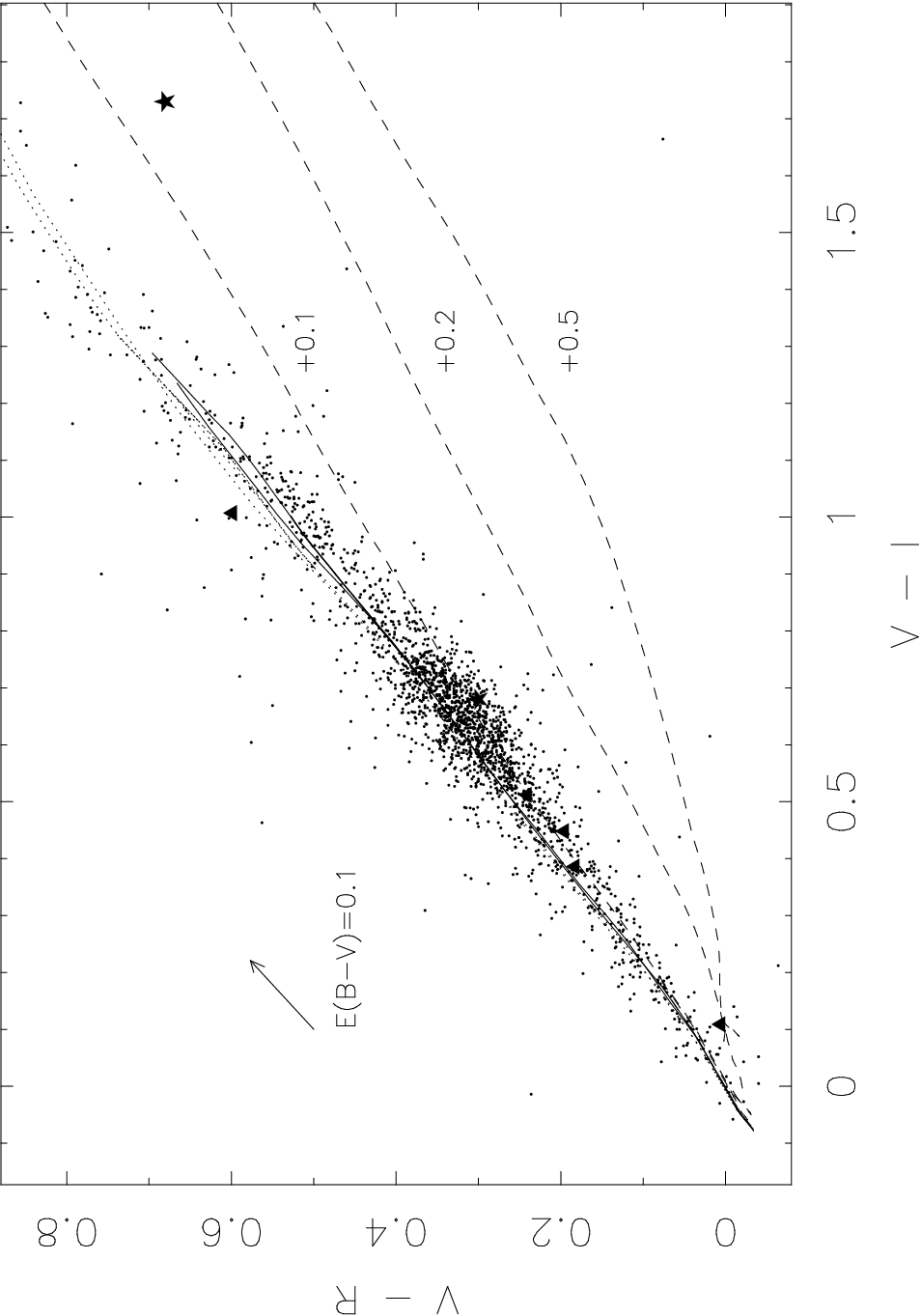}
\end{figure}

\begin{figure}[]
\plotone{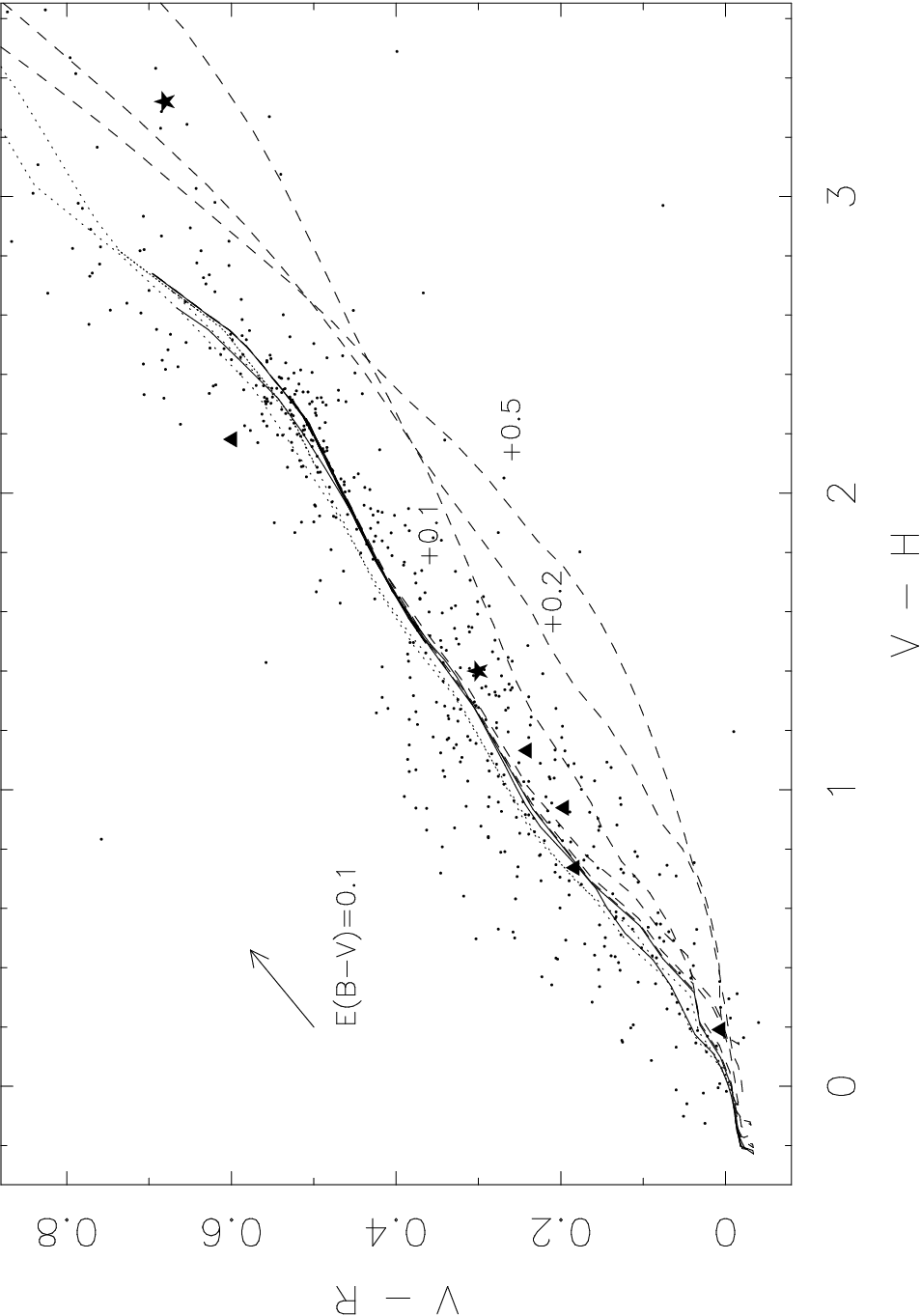}
\end{figure}

\begin{figure}[]
\plotone{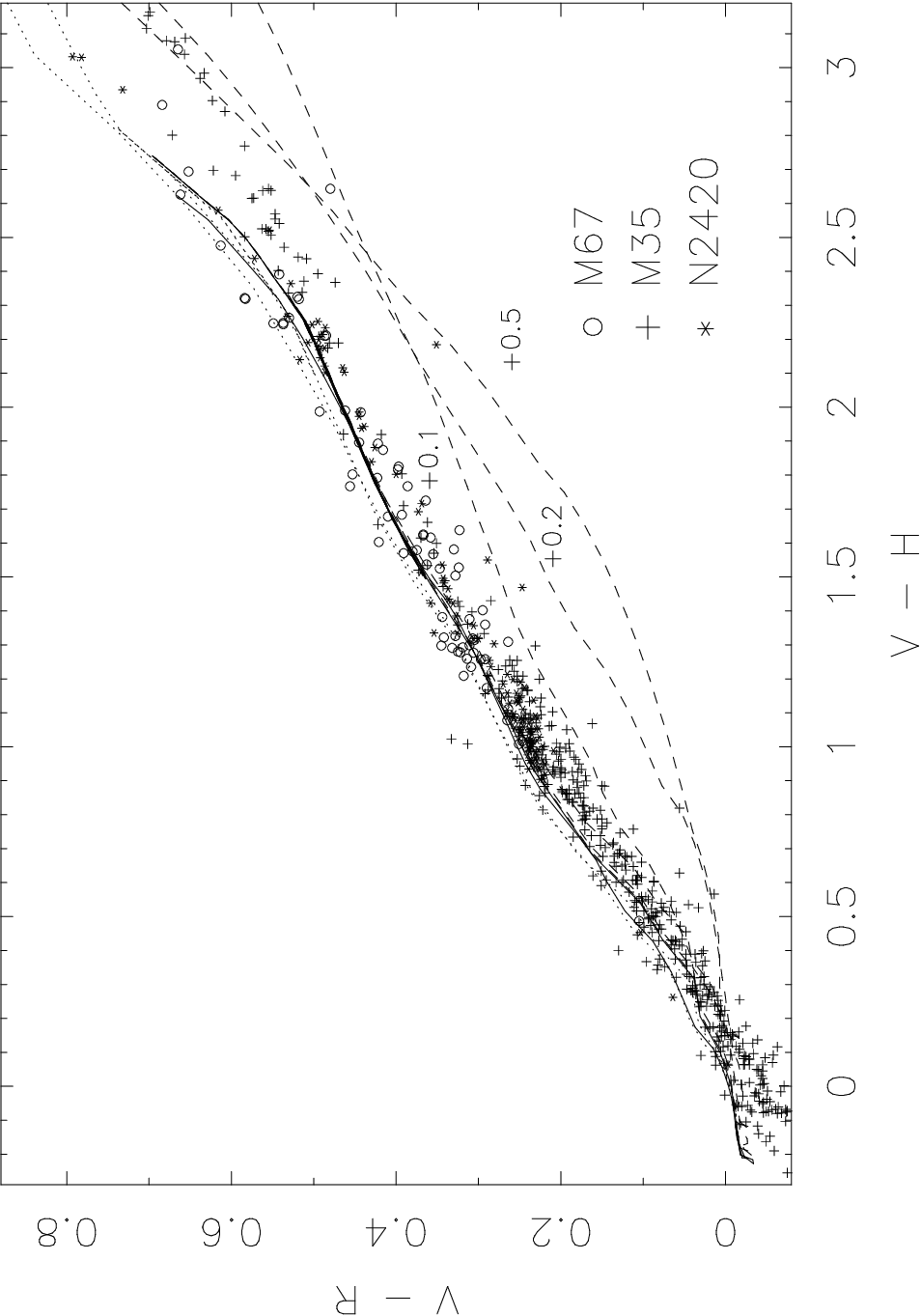}
\end{figure}

\begin{figure}[]
\plotone{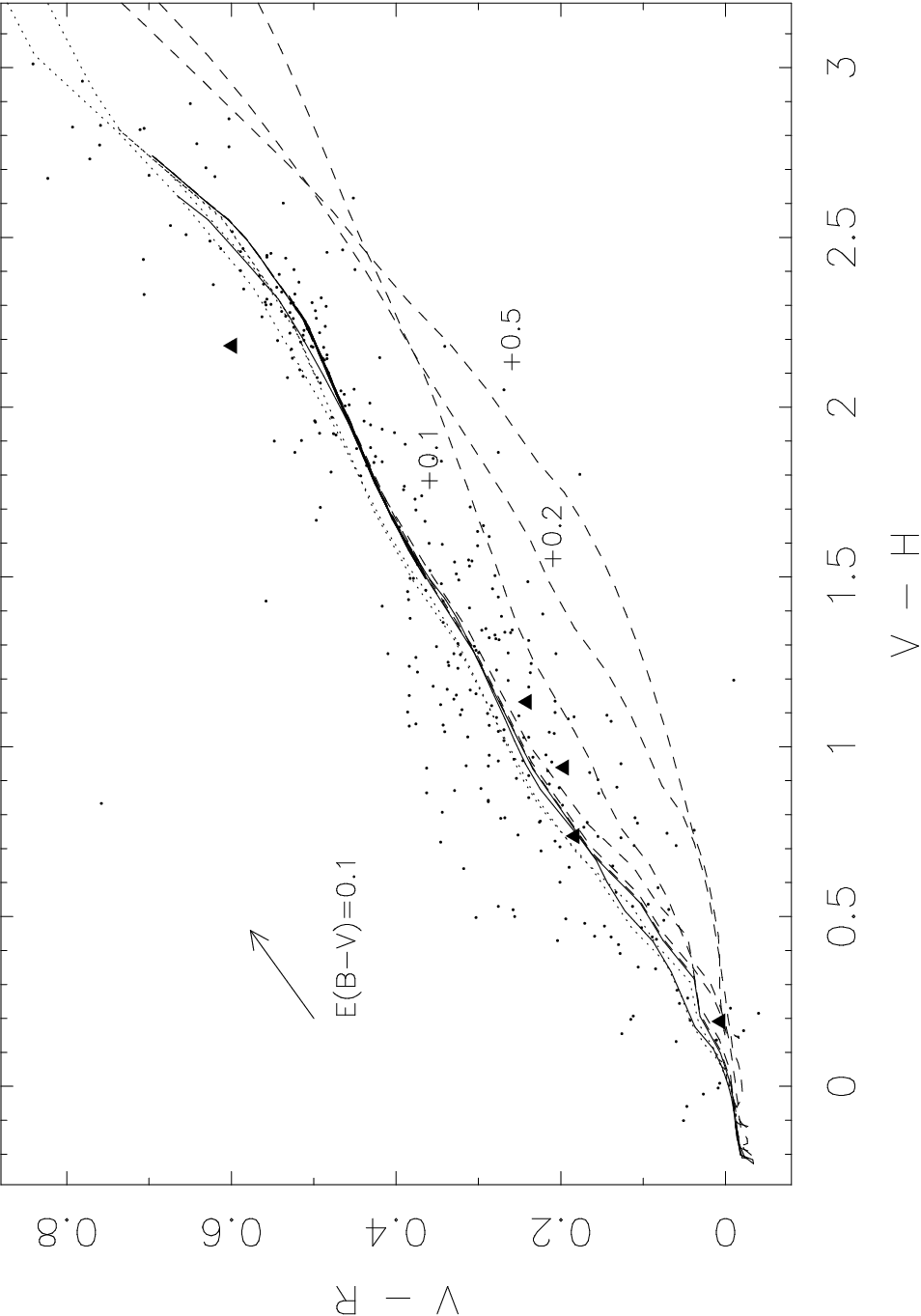}
\end{figure}

\begin{figure}[]
\plotone{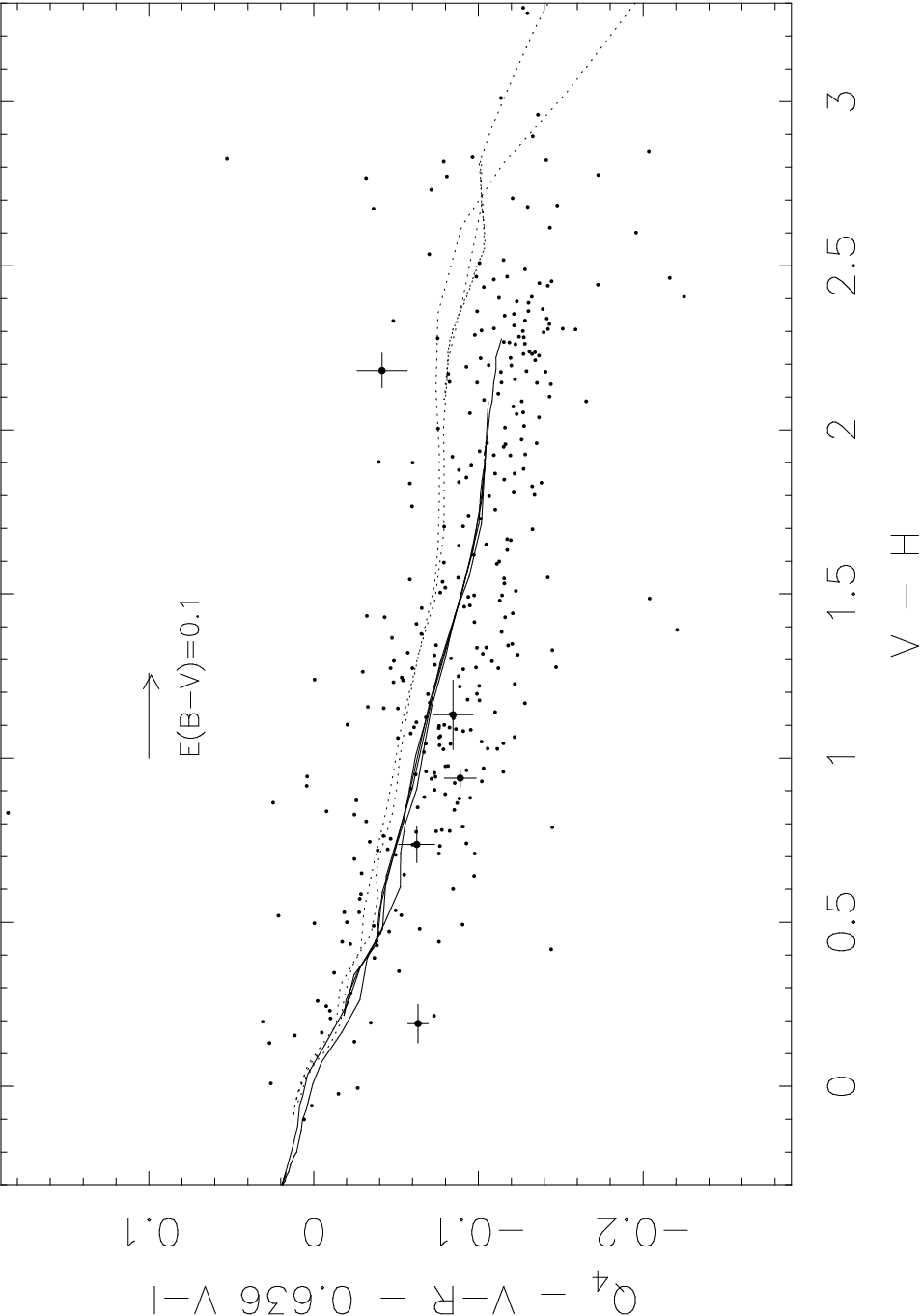}
\end{figure}

\begin{figure}[]
\plotone{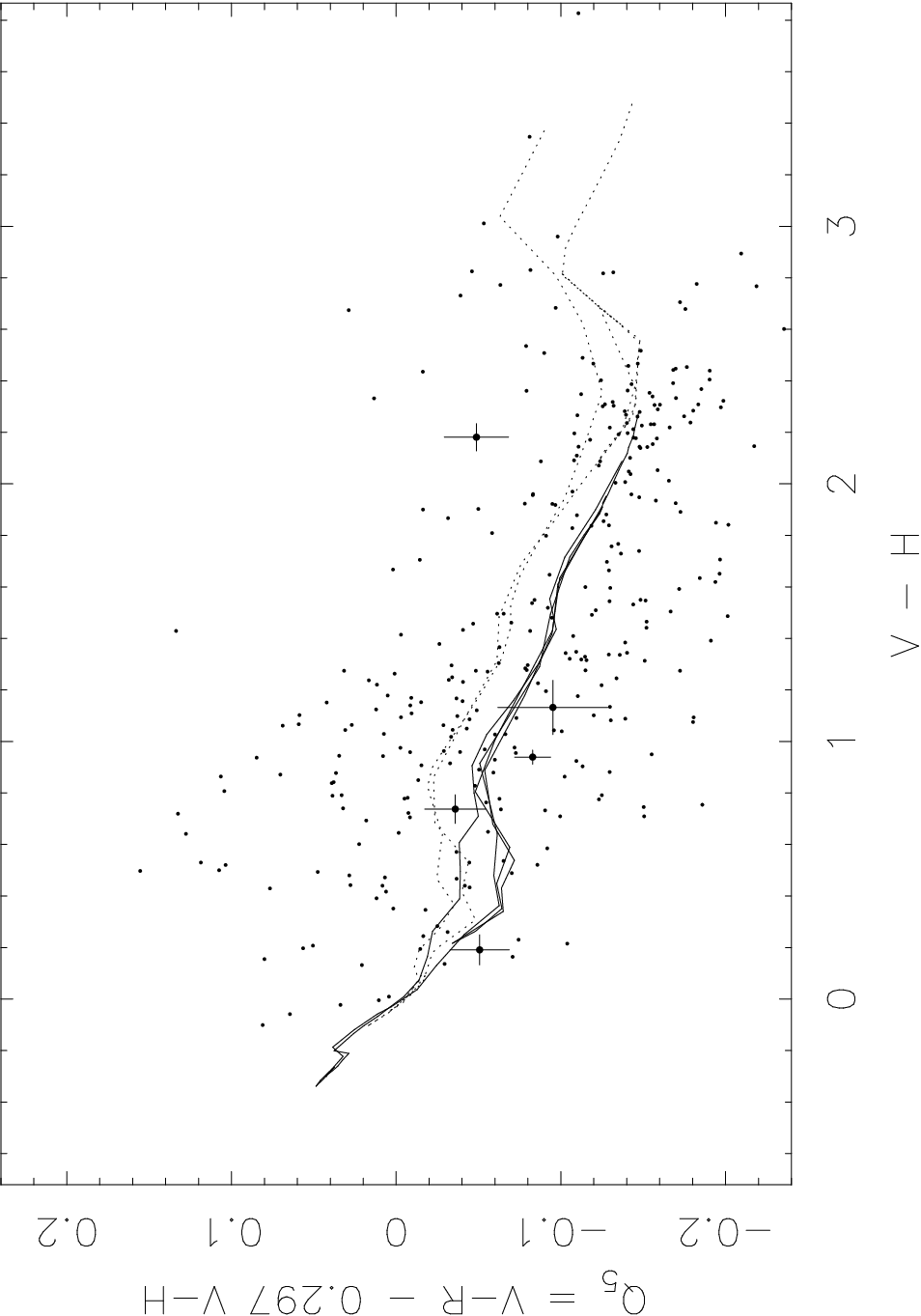}
\end{figure}

\begin{figure}[]
\plotone{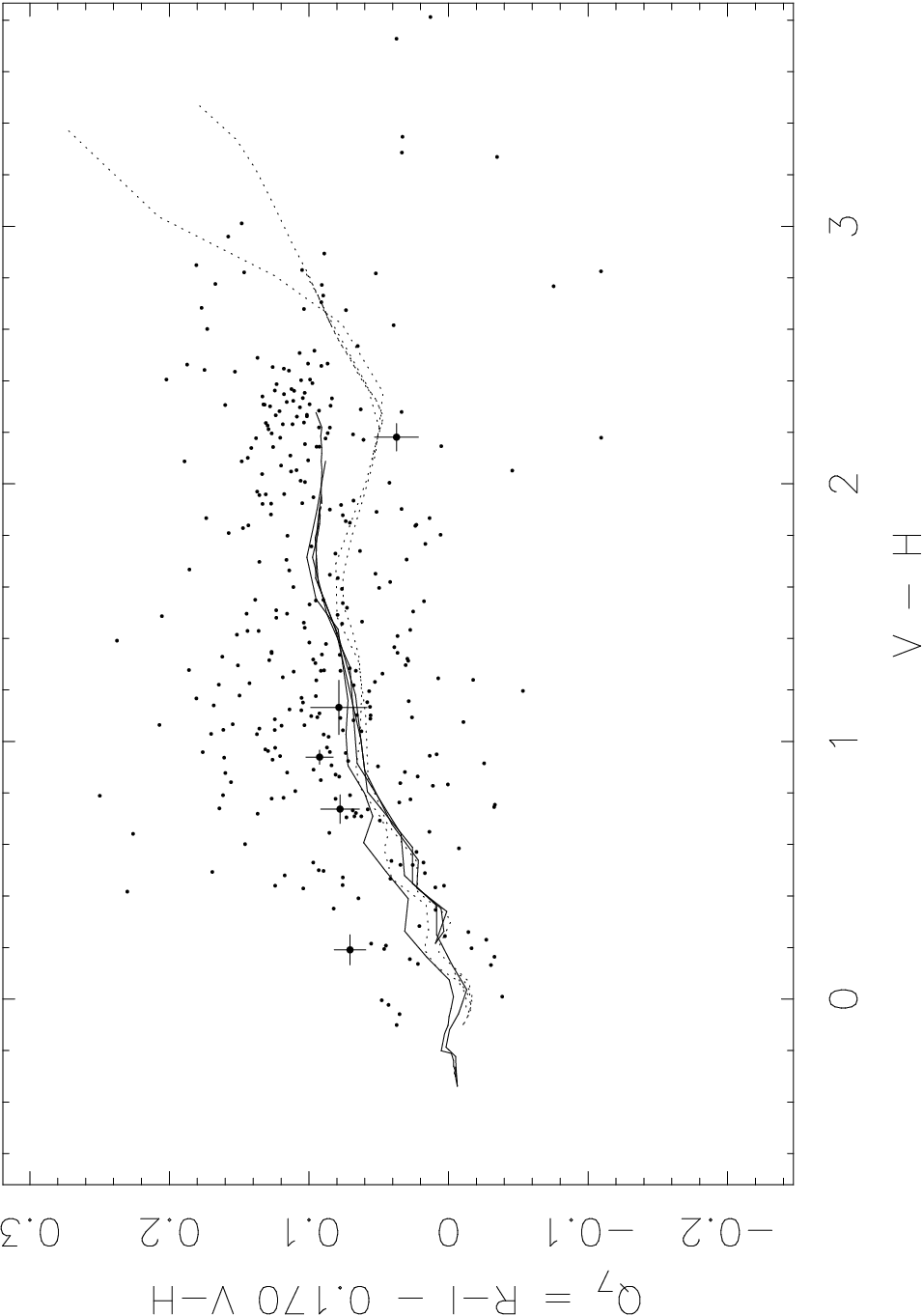}
\end{figure}

\begin{figure}[]
\plotone{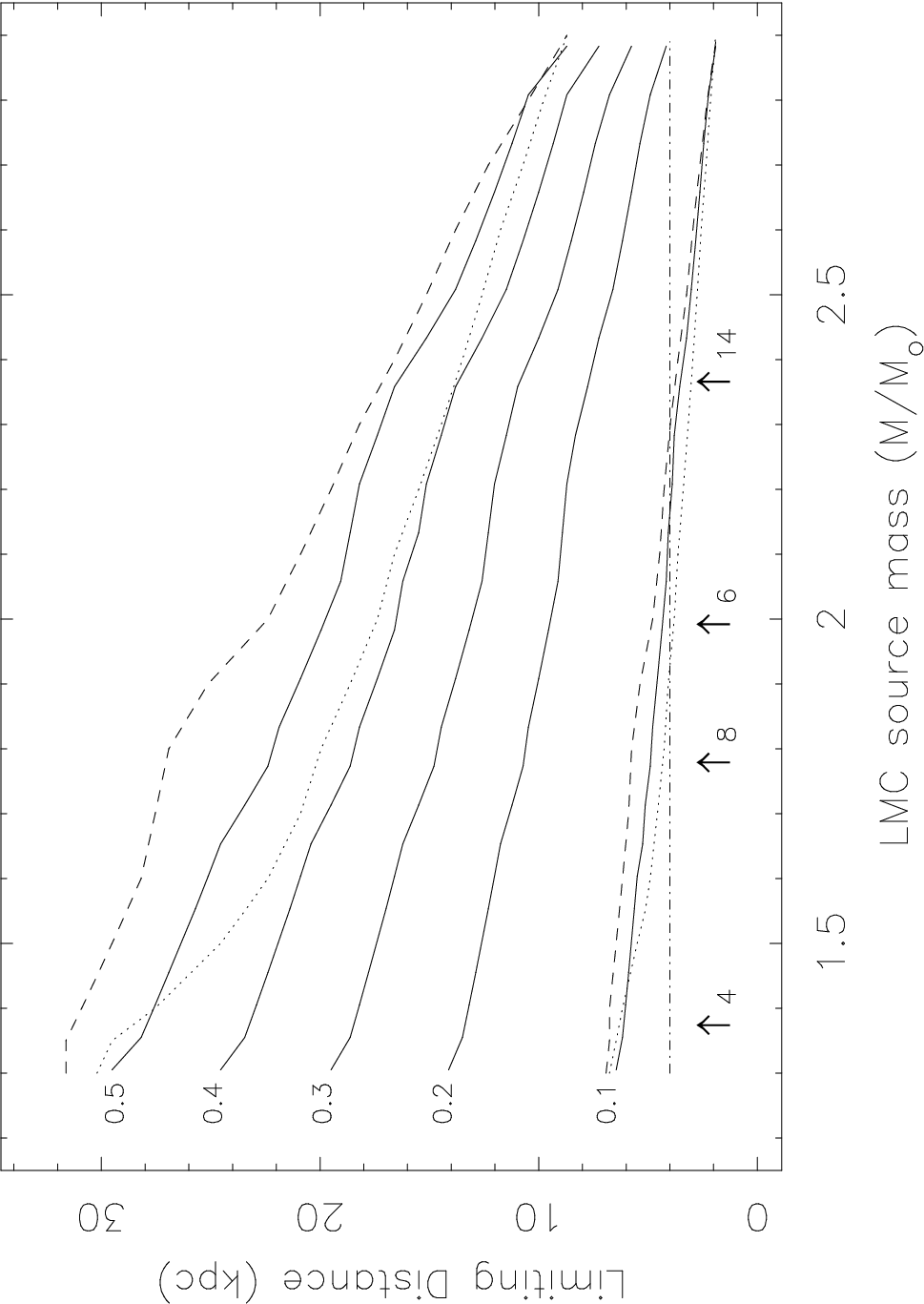}
\end{figure}

\end{document}